\documentclass[aps,prx,reprint,superscriptaddress,longbibliography,nofootinbib]{revtex4-2}

\usepackage[T1]{fontenc}
\usepackage[utf8]{inputenc}
\usepackage{amsmath,amssymb,amsfonts}
\usepackage{bm}
\usepackage{graphicx}
\usepackage{xcolor}
\usepackage{booktabs}
\usepackage{physics}
\usepackage[colorlinks=true, citecolor=blue, linkcolor=blue, urlcolor=blue]{hyperref}

\newcommand{\SU}{\mathrm{SU}}
\newcommand{\su}{\mathfrak{su}}

\begin{document}

\title{Lie Group Diffusion Models for Hardware-Aware Quantum Circuit Synthesis}

\author{Jyotirmai Singh}
\email{joesingh@stanford.edu}
\affiliation{ Department of Physics,  Stanford University}

\date{\today}

\begin{abstract}
An important task in quantum computing is unitary circuit synthesis compatible
with physical hardware constraints. This problem has a natural hybrid structure as
local single-qubit gates are continuous variables on the Lie group $\SU(2)$ while the 
entangling circuit structure is discrete and hardware-dependent. In this work, we use 
generative models to perform quantum circuit synthesis incorporating
both the natural $\SU(2)$ manifold geometry of quantum gates and
hardware constraints that determine the overall circuit structure. Our model comprises 
two components: a circuit skeleton selector that chooses an entangling circuit and a diffusion model that generates quantum gates on the given circuit template by performing diffusion on  the curved manifold $\SU(2) \simeq S^3$ itself. We demonstrate this approach with unitary
compilation of physically motivated three-qubit Hamiltonian simulation targets such as the Transverse Field
Ising Model and the Heisenberg-XXZ Model and show that Lie group diffusion outperforms comparable baselines. The synthesised circuits can also be customised subject to constraints,
which we demonstrate by producing circuits with large and small gate rotation angles for
the same target unitary evolution. We also investigate the fidelity-complexity frontier
of the synthesised gates to demonstrate that the circuit selector learns to select circuits that balance fidelity with complexity rather than collapsing onto
the most expansive entangling template. These results demonstrate that Lie group diffusion provides a natural generative framework for hardware-aware quantum circuit synthesis.
\end{abstract}

\maketitle

\section{Introduction}
Quantum computing is poised to become a transformative technology, with potential impact across fundamental science \cite{feynman1982simulating, RevModPhys.86.153, cao2019quantum, preskill2019simulating}, optimisation \cite{farhi2014quantum}, and machine learning, including provable advantages in learning from quantum experiments \cite{huang2025generative} and massive classical datasets \cite{zhao2026exponential}.

Realising these applications requires compiling desired quantum operations into circuits that can be executed on a target device. In general, quantum circuit synthesis asks for a circuit $C$ that implements a target unitary operation \(U_{\star}\), subject to hardware constraints. These constraints include the native gate set, the connectivity graph of the quantum processor unit (QPU), and the different error rates associated with one and two-qubit gates. The synthesis problem thus has a natural mixed structure: the entangling pattern is discrete and architecture-dependent, while the local single-qubit gates that populate a circuit template are continuous variables on the compact Lie group \(\SU(2)\).

Machine learning methods have recently shown significant promise for quantum circuit design, including neural-network predictors \cite{zhang2021neural}, reinforcement learning \cite{zhang2020topological, ruiz2025alphatensor}, and differentiable search \cite{zhang2022differentiable}. Diffusion models \cite{sohldickstein2015deep} provide another powerful generative approach and have also been applied to quantum circuit synthesis \cite{furrutter2024quantum}. Existing diffusion-based approaches, however, typically represent circuits using Euclidean tensor representations and perform denoising in that ambient space.

Here we take a complementary approach that builds the geometry and hardware constraints of the synthesis problem directly into the generative model. We introduce a circuit-synthesis architecture consisting of two coupled components: a learned circuit skeleton selector and a skeleton-conditioned diffusion prior based on a denoising diffusion probabilistic model (DDPM) \cite{ho2020denoising}. The selector chooses a discrete circuit template, with training labels determined by both target fidelity and a hardware-aware cost that depends on qubit connectivity and gate error rates. Conditional on this template, the diffusion model generates continuous local one-qubit gates, which undergo local optimisation/refinement on the same $\SU(2)$ manifold to obtain the final circuit. Unlike Euclidean denoising approaches, the diffusion process is performed directly on the curved manifold \cite{debortoli2022riemannian, huang2022riemannian} using the Lie-group structure of $\SU(2)$. The resulting model treats circuit synthesis as a hardware-constrained physical design problem: it selects an allowed entangling skeleton and then generates continuous gate parameters on the correct geometric manifold.

We demonstrate this approach on three-qubit Hamiltonian simulation targets using a nearest-neighbor CZ circuit skeleton architecture. Across five Hamiltonian target families including physically motivated targets such as the Transverse-Field Ising Model (TFIM) and Heisenberg-XXZ Model, the diffusion prior improves final synthesis success relative to both a Haar-random prior and a stronger Clifford-group aware prior. Ablation experiments show that diffusion and local refinement play complementary roles: the diffusion model acts as a broad, geometry-aware proposal prior while local manifold refinement performs the final high-fidelity optimisation.

Beyond raw compilation accuracy, we show that the diffusion prior can be steered toward different styles of successful circuits. We demonstrate this by
biasing the model towards synthesised circuits with higher or lower gate
rotation angles for a given target unitary, showing that it
can be steered based on physical constraints. Finally, by evaluating each target across the full circuit skeleton library using the diffusion prior, we show that the circuit selector learns a hardware-aware fidelity--complexity tradeoff, concentrating probability on compact circuit templates with lower hardware overhead when they are sufficient and selecting deeper entangling templates only when required. Together, these results show that Lie-group diffusion performs an effective geometry-aware generative prior for hardware-constrained quantum circuit design.

\section{Diffusion on $\SU(2)^n$ Circuit Gates}
\label{sec:diffusion-math}

\subsection{Quaternion Representation of Single Qubit Gates}
A local one-qubit operation can in general be expressed as an 
element of the Lie group $\SU(2)$. For the purposes of computational ease and efficiency, we use the quaternion representation of
$\SU(2)$. In this, a one-qubit gate is represented by a unit 
quaternion:

\begin{equation}
    \boldsymbol{q} = (w, x, y, z), \quad \|\boldsymbol{q}\|^2 = 1
    \label{eq:quaternion-def}
\end{equation}

This is mapped to an element $U \in \SU(2)$ by

\begin{equation}
    U(\boldsymbol{q}) = \begin{pmatrix}
                        w - iz & - y - ix \\
                        y - ix & w + iz
                    \end{pmatrix}
                    = w I -i(x\sigma_x + y\sigma_y + z\sigma_z)
    \label{eq:quaternion-su2-conversion}
\end{equation}

where $I$ is the identity and $\sigma_x$, $\sigma_y$, $\sigma_z$ are the Pauli matrices

\begin{equation*}
        \sigma_x = \begin{pmatrix}
                        0 & 1 \\
                        1 & 0
                    \end{pmatrix}, \quad 
        \sigma_y = \begin{pmatrix}
                        0 & -i \\
                        i & 0
                    \end{pmatrix}, \quad
        \sigma_z = \begin{pmatrix}
                        1 & 0 \\
                        0 & -1
                    \end{pmatrix}
\end{equation*}

A single qubit operation $U$ that rotates the state around the axis 
$\boldsymbol{n} = (n_x, n_y, n_z)$ by an angle $\theta$ can be represented 
using the exponential map from the Lie algebra $\su(2) = \mathrm{span}_{\mathbb{R}}(i\sigma_x, i\sigma_y, i\sigma_z)$ to the 
Lie group $\SU(2)$

\begin{equation}
    U(\boldsymbol{q}) = \exp\left[-i\hat{\boldsymbol{n}
    }\cdot \boldsymbol{\sigma}\frac{\theta}{2}\right]
    \label{eq:exponential-map-su2}
\end{equation}

when $w = \cos\left(\frac{\theta}{2}\right)$ and $(x, y, z) = \hat{\boldsymbol{n}}\sin\left(\frac{\theta}{2}\right)$. Since
$\su(2) \simeq \mathbb{R}^3$ as vector spaces, we identify 
the element $i(x\sigma_x + y\sigma_y + z\sigma_z) \in \su(2)$ with
the real vector $\boldsymbol{v} = (x, y, z) \in \mathbb{R}^3$ and 
define the matrix exponential as

\begin{equation}
    \exp(\boldsymbol{v})=
\begin{cases}
\left(\cos\|\boldsymbol{v}\|,\dfrac{\sin\|\boldsymbol{v}\|}{\|\boldsymbol{v}\|}\boldsymbol{v}\right), & \|\boldsymbol{v}\|\neq 0, \\[1.2em]
(1,0,0,0), & \|\boldsymbol{v}\|=0.
\end{cases}
\label{eq:exp-defn}
\end{equation}

Thus overall, elements of $\SU(2)$, represented with unit quaternions,
are generated by exponentiating vectors in the Lie algebra $\su(2)$,
which are equivalently identified with vectors in 
$\mathbb{R}^3$. 

We can also define a logarithm in the reverse direction to obtain
a Lie algebra vector that exponentiates to the desired unitary
operator. Let
$\boldsymbol{q} = (w, x, y, z) = (w, \boldsymbol{u})$ be a quaternion identified as an 
element of $\SU(2)$. Define $\phi = \frac{\theta}{2} =  \arctan2(\|\boldsymbol{u}\|, w)$.
Then the principal logarithm is
\begin{equation}
\log(q)
=
\begin{cases}
\dfrac{\mathbf u}{\|\mathbf u\|} \phi, & \|\mathbf u\|>0,\\[1ex]
0, & \|\mathbf u\|=0.
\end{cases}
\label{eq:log-defn}
\end{equation}
This returns the Lie algebra vector \(v\in\mathbb R^3\) whose exponential
is $q$ on the chosen principal branch.

\begin{figure*}
    \centering
    \includegraphics[width=\linewidth,trim=10pt 10pt 10pt 10pt,clip]{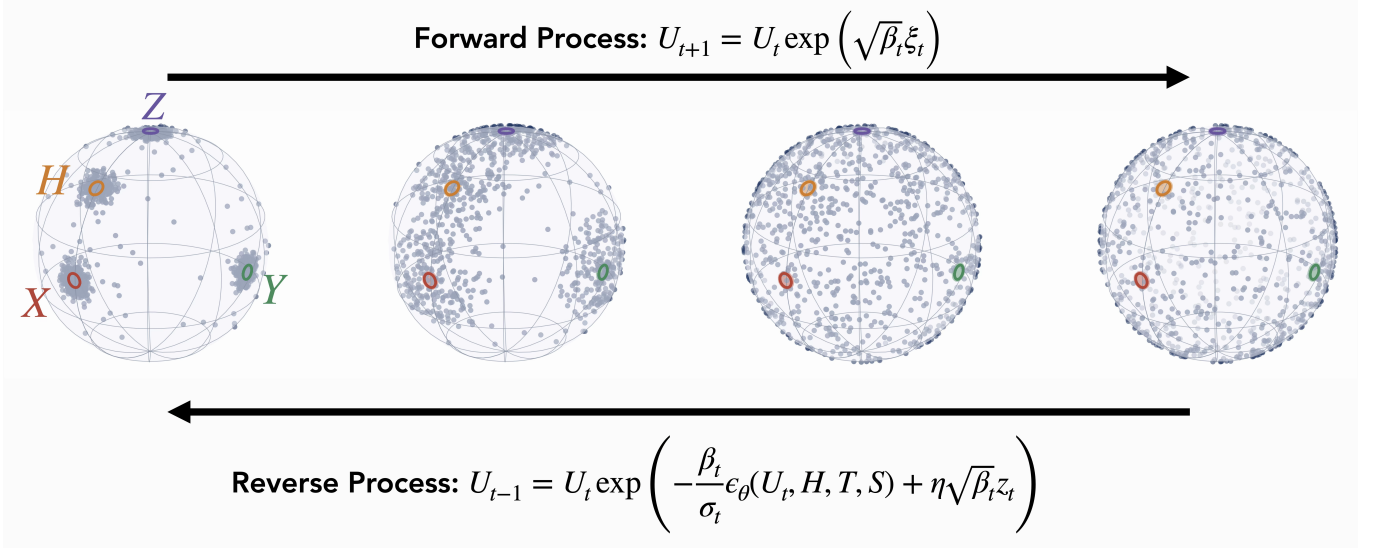}
    \caption{
Schematic of heat-kernel diffusion for single-qubit gates.
Local gates are represented as elements of $\SU(2)$ and visualised here through their action on the Bloch sphere.
In the forward process, clean gate distributions are progressively corrupted by right-multiplicative Lie-algebra noise,
\(U_{t+1}=U_t\exp(\sqrt{\beta_t}\xi_t)\), spreading structured clusters toward a Haar-like distribution.
In the reverse process, a learned denoiser predicts tangent-space updates that transport noisy gates back toward the data distribution. 
The denoising target incorporates the $\SU(2)$ heat kernel
so that the learned denoiser learns about the
nonlinear geometry of the manifold.
The Bloch-sphere panels are illustrative: the implemented diffusion process evolves on $\SU(2)\simeq S^3$, and circuit synthesis applies this update independently across local-gate slots in $\SU(2)^n$.
}
    \label{fig:bloch-sphere-diffusion}
\end{figure*}

\subsection{Forward Diffusion Process}
As in standard diffusion models, the forward process gradually corrupts
a clean datapoint with noise. Here the datapoint is a single-qubit
unitary \(U_t \in \mathrm{SU}(2)\), so the corruption process must remain
on the group manifold. Using the identification
\(\mathfrak{su}(2) \simeq \mathbb{R}^3\) defined above, we sample a
Gaussian tangent increment and map it back to the group with the
exponential map:
\begin{equation}
U_{t+1}
=
U_t \exp\!\left(\sqrt{\beta_t}\,\xi_t\right),
\quad
\xi_t \sim \mathcal{N}(0,I_3)
\label{eq:forward-process-single}
\end{equation}
Here \(\xi_t \in \mathbb{R}^3\) is interpreted as a Lie-algebra tangent
vector. The scalar
\(\beta_t\) is the noise schedule that sets the
standard deviation of the tangent-space Brownian increment, analogous to the noise schedule in Euclidean DDPMs.

For a quantum circuit with $n$ single-qubit local-gate slots, forward
noising is applied independently to each slot:
\begin{equation}
U^{(j)}_{t+1}
=
U^{(j)}_t \exp\!\left(\sqrt{\beta_t}\,\xi^{(j)}_t\right),
\quad
j = 1, \dots, n
\label{eq:forward-process-multiple}
\end{equation}

\subsection{Heat Kernel Denoising}
\subsubsection{Single-Qubit Denoising Process}
During the reverse process, a neural-network is trained to predict
the noise so as to undo its action. Like the forward noising process,
the reverse denoising process must also respect the nonlinear geometry
of $\SU(2) \simeq S^3$.

Given the initial datapoint $U_0$, the full forward process is a
cumulative product of Lie algebra perturbations:

\begin{equation}
    U_t = U_0 \prod_{s=1}^t \exp\left(\sqrt{\beta_s}\xi_s\right), \quad \xi_s \sim \mathcal{N}(0, I_3)
    \label{eq:full-forward-process}
\end{equation}

Note that due to the noncommutativity of $\SU(2)$, the time-ordering of the product matters and must be preserved. Eq.~\ref{eq:full-forward-process} defines a conditional density on the group 

\begin{equation}
    q_t(U_t | U_0) = K_{\sigma_t^2} (U_0^{-1} U_t) 
\end{equation}

where $K_{\sigma_t^2}$ is the heat kernel on $\SU(2)$ and
$\sigma_t^2 = \sum_{s=1}^{t} \beta_s$ is the cumulative variance from timestep 0 to $t$. In the local limit with small $\sigma_t$ and  $U_0^{-1}U_t \approx I$, the $\SU(2)$ heat kernel reduces to a standard 
Gaussian and we recover Euclidean diffusion.

The network is then trained to predict the target, which is
the score $\nabla \log q_t$ scaled by the cumulative standard deviation:

\begin{equation}
    \epsilon^{\rm target}_t = -\sigma_t \nabla\log q_t
    \label{eq:simple-target}
\end{equation}

Let \(R_t = U_0^{-1}U_t\) be the relative displacement on the manifold
caused by the forward process. Define
\(\xi_t^{\rm rel}=\log R_t\) to be the Lie-algebra vector that
generates this displacement, and let $\phi_t = \|\xi_t^{\rm rel}\|$
be the \(\mathrm{SU}(2)\) geodesic displacement angle, equal to
half of the corresponding physical single-qubit rotation angle.

The heat kernel on $\mathrm{SU}(2)$ is conjugation-invariant and therefore radial: it depends on \(R_t=U_0^{-1}U_t\) only through the displacement angle $\phi_t$. This is reflected in the spectral expansion \(K_{\sigma_t^2}(\phi)\), which contains no dependence on the rotation axis \cite{fegan1983fundamental, bismut2008hypoelliptic}:

\begin{equation}
    K_{\sigma_t^2}(\phi_t) \propto \sum_{m=1}^{\infty} m e^{-\frac{1}{2}(m^2 - 1)\sigma_t^2}\frac{\sin(m\phi_t)}{\sin\phi_t} 
    \label{eq:heat-kernel-spectral-expn}
\end{equation}

Using the chain rule, the score function becomes

\begin{equation}
    \nabla \log q_t = \partial_{\phi_t} \log(K_{\sigma_t^2}(\phi_t)) \frac{\xi_t^{\rm rel}}{\phi_t}
    \label{eq:score-function}
\end{equation}

Here, the unit vector $\frac{\xi_t^{\rm rel}}{\phi_t}$ is the 
direction of the score tangent vector while the scalar derivative 
encodes how strongly the density changes. The full noise target is
obtained then from Eq.~\ref{eq:simple-target}:

\begin{equation}
    \epsilon^{\rm target}_t = -\sigma_t \partial_{\phi_t} \log\left(K_{\sigma_t^2}(\phi_t)\right) \frac{\xi_t^{\rm rel}}{\phi_t}
    \label{eq:noise-target}
\end{equation}

For numerical stability, at small diffusion times $t$ we use the 
local approximation

\begin{equation}
    \partial_{\phi}\log\left(K_{\sigma_t^2}(\phi)\right) \approx \frac{1}{\phi} - \cot \phi - \frac{\phi}{\sigma_t^2} 
\end{equation}

while near the origin, we use the small $\phi$ approximation

\begin{equation}
    \frac{1}{\phi} - \cot \phi \approx \frac{\phi}{3} + \frac{\phi^3}{45} + \frac{2\phi^5}{945} + \mathcal{O}(\phi^7)
\end{equation}

Finally, the loss function is given by the standard MSE error between
the prediction and Eq.~\ref{eq:noise-target}:

\begin{equation}
    \mathcal{L} = \mathbb{E}\left[\left|\left|\epsilon^{\theta}_t - \epsilon_{t}^{\rm target}\right|\right|^2\right]
\end{equation}

\subsubsection{Circuit Denoising on $n$ Single-Qubit Slots}
For a circuit with \(n\) local single-qubit gate slots, the diffusion
state is a point on the product manifold
\begin{equation}
\mathcal M = \mathrm{SU}(2)^n .
\end{equation}
We apply the same heat-kernel construction independently to each slot.
Writing
\begin{equation}
\mathbf U_t=(U_t^{(1)},\ldots,U_t^{(n)}),
\end{equation}
the relative displacement for slot \(j\) is
\begin{equation}
R_{t,j} = (U_0^{(j)})^{-1}U_t^{(j)},\quad
\xi_{t,j}^{\rm rel}=\log R_{t,j},\quad
\phi_{t,j}=\|\xi_{t,j}^{\rm rel}\|.
\end{equation}
Since the forward noising process factorises over slots, the conditional
density is
\begin{equation}
q_t(\mathbf U_t\mid \mathbf U_0)
=
\prod_{j=1}^{n}
K_{\sigma_t^2}(\phi_{t,j})
\end{equation}
Thus the epsilon target also factorises slotwise:
\begin{equation}
\epsilon^{\rm target}_{t,j}
=
-\sigma_t
\partial_\phi \log K_{\sigma_t^2}(\phi_{t,j})
\frac{\xi^{\rm rel}_{t,j}}{\phi_{t,j}},
\quad
j=1,\ldots,n
\label{eq:multi-noise-target}
\end{equation}
The denoising network therefore predicts an \(n\times 3\) tangent-vector
field, with one three-dimensional Lie-algebra vector for each local gate
slot. Applying reverse updates through the exponential map keeps every
slot on \(\mathrm{SU}(2)\), and hence keeps the full circuit state on
\(\mathrm{SU}(2)^n\)

\begin{figure*}
    \centering
    \includegraphics[width=1.0\linewidth]{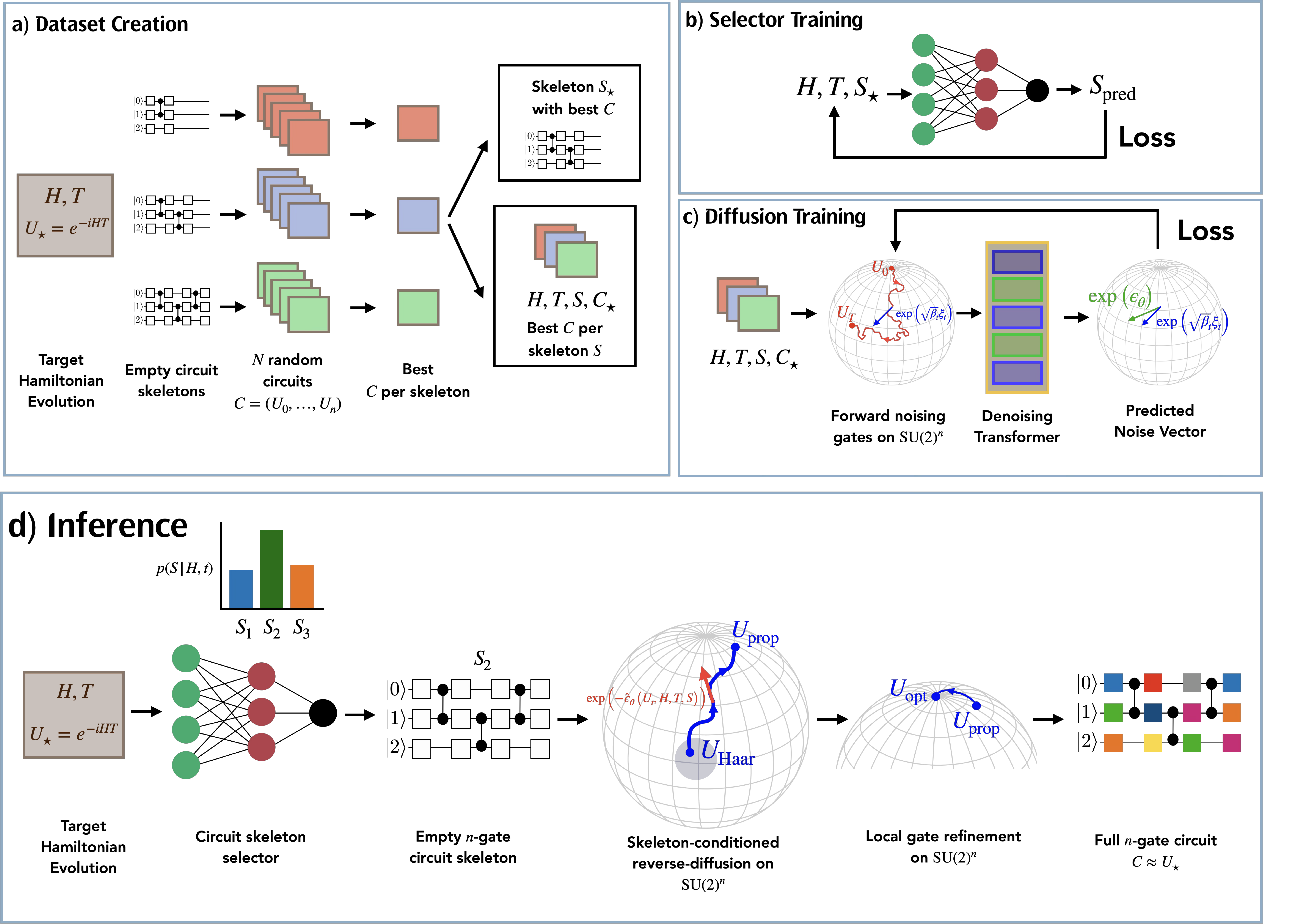}
    \caption{
Overview of the skeleton-conditioned circuit synthesis pipeline.
(a) Synthetic training data is constructed by pairing target Hamiltonian evolutions
$U_\star=\exp(-iHT)$ with candidate circuits on a fixed library of CZ skeletons.
For each target and skeleton, random local-gate configurations are generated,
ranked by fidelity, locally refined, and reduced to the best refined circuit for
that skeleton. Comparing the refined circuits across skeletons gives the
cost-aware skeleton label used for selector training.
(b) The skeleton selector is trained as a supervised classifier mapping target
features $(H,T)$ to a skeleton label $S_\star$.
(c) The diffusion model is trained on circuits conditioned on
$(H,T,S)$: clean local gates are noised on the product manifold $SU(2)^n$, and a
skeleton-conditioned denoising transformer predicts the heat-kernel noise target.
(d) At inference time, a new target Hamiltonian is passed through the selector,
the chosen empty skeleton is filled by reverse diffusion on $SU(2)^n$, and the
resulting proposal is locally refined to obtain the final circuit
$C \simeq U_\star$.
}
    \label{fig:main-figure}
\end{figure*}

\subsection{Reverse Sampler on \texorpdfstring{$\SU(2)^n$}{SU(2)	extasciicircum n}}
Given a noisy circuit state
\begin{equation}
U_t = \left(U_t^{(1)},\ldots,U_t^{(n)}\right)\in SU(2)^n,
\end{equation}
the denoiser predicts one tangent vector per slot,
\begin{equation}
\epsilon_{\theta,t,j}
=
\epsilon_\theta(U_t,H, T,S)_j
\in \mathbb{R}^3.
\end{equation}
Where $H, T, S$ refer to the target Hamiltonian, evolution time, and circuit skeleton respectively.
Since the target in Eq.~\ref{eq:multi-noise-target} is
\begin{equation}
\epsilon^{\mathrm{target}}_{t,j}
=
-\sigma_t \nabla_{U_t^{(j)}} \log q_t,
\end{equation}
we convert the predicted epsilon back into a score estimate by
\begin{equation}
\widehat{s}_{t,j}
=
-\frac{1}{\sigma_t}\epsilon_{\theta,t,j}.
\end{equation}
We then use a right-invariant Euler--Maruyama reverse step on the group:
\begin{align}
\Delta_{t,j} &= \beta_t \widehat{s}_{t,j}
+
\eta \sqrt{\beta_t} z_{t,j} \\
&=
-\frac{\beta_t}{\sigma_t}\epsilon_{\theta,t,j}
+
\eta \sqrt{\beta_t} z_{t,j},
\quad
z_{t,j}\sim \mathcal{N}(0,I_3),
\end{align}
and update each local gate by exponentiating this tangent increment:
\begin{equation}
U_{t-1}^{(j)}
=
U_t^{(j)}
\exp(\Delta_{t,j}).
\end{equation}
At the final step we set the stochastic term to zero. The parameter
\(\eta\) controls sampler stochasticity, with \(\eta=0\) giving a
deterministic reverse trajectory.

\section{Hamiltonian-Conditioned Circuit Synthesis Model}

The model's goal is to generate a quantum circuit $C$ that implements the target unitary $U_{\star} = \exp(-iHT)$
where $H$ is the target Hamiltonian to be applied for a time $T$. The circuit $C$
is composed of gates $U_1 \dots, U_n$ that are sequenced on a particular circuit
skeleton $S$. The model does this through two separate parts: 1. the selector model
that picks the best skeleton $S$ conditioned on $H$ and $T$ and 2. the diffusion
model that produces the gates $U_1, \dots , U_n$ conditioned on $H, T, S$. Both
of these are trained separately but using the same training dataset.

\subsection{Circuit Templates}
We focus on three-qubit targets because two-qubit unitary synthesis is largely analytically controlled by the Cartan/KAK decomposition \cite{Vidal2005Universal}, whereas three-qubit unitaries already inhabit a substantially richer geometry for which no comparably compact canonical entangling parametrisation is available. While analytic decompositions exist for special three-qubit gate
classes and provide worst-case CNOT upper bounds \cite{pawlowski2025real3qubittriality}, our goal is different: to learn
shallow CZ-skeleton decompositions of structured Hamiltonian evolutions with
continuous local \(\mathrm{SU}(2)\) gates.

In this work we restrict attention to a fixed library of three-qubit CZ 
circuit skeletons which encode a simple nearest-neighbor three-qubit hardware graph. More generally, the same construction could encode the connectivity graph and native entangling gates of a specific QPU architecture. The skeleton specifies the entangling pattern with its
distribution of CZ gates, and all empty slots in the skeleton are filled with 
one-qubit gates sampled from $\SU(2)$. For a skeleton with $k$ CZ gates, there
are $k+1$ local gate layers, corresponding to $3k+3$ total local gates in a 
three-qubit configuration. We allow $k$ to range from 0 to 5. Table \ref{tab:skeleton-library} outlines all the skeleton templates schematically, while Fig.~\ref{fig:main-figure}a) illustrates the skeletons 1-CZ 01, 
2-CZ, and 3-CZ pictorially. The 4 and 5-CZ gate templates
are constructed analogously. Unlike the higher templates which entangle all qubits by default, the 1-CZ templates
leave one qubit unentangled so we incorporate two 1-CZ templates to test whether the model can learn where to put entanglement between qubits.

\begin{table}[t]
\centering
\begin{tabular}{lccc}
\hline
Skeleton & CZ edge sequence & $N_{\rm CZ}$ & $N_{\rm local}$ \\
\hline
0-CZ    & -- & 0 & 3  \\
1-CZ 01  & $(0,1)$ & 1 & 6  \\
1-CZ 12  & $(1,2)$ & 1 & 6  \\
2-CZ   & $(0,1),(1,2)$ & 2 & 9  \\
3-CZ   & $(0,1),(1,2),(0,1)$ & 3 & 12 \\
4-CZ     & $(0,1),(1,2),(0,1),(1,2)$ & 4 & 15 \\
5-CZ   & $(0,1),(1,2),(0,1),(1,2),(0,1)$ & 5 & 18 \\
\hline
\end{tabular}
\caption{
Three-qubit CZ skeleton library used by the selector.
Each skeleton is specified by its ordered CZ edge sequence on the line
$q_0$--$q_1$--$q_2$. For a skeleton with $k$ CZ gates, arbitrary local
$SU(2)$ gates are placed in $k+1$ layers, giving $3(k+1)$ local-gate slots.
}
\label{tab:skeleton-library}
\end{table}

\subsection{Hardware-Aware Circuit Score}

Given a target unitary $U_{\star}$ and a candidate circuit $C = (U_1,\dots, U_n)$ with skeleton $S$, we define two metrics that are used for evaluation. First, for a circuit $C$ with target $U_{\star}$, we quantify circuit accuracy using the process fidelity between the target
unitary \(U_{\star}=\exp(-iHT)\) and the synthesised circuit unitary \(C\) as,
\begin{equation}
F(C, U_{\star})
=
\frac{1}{d^2}
\|\operatorname{Tr}(U_{\star}^\dagger C)\|^2
\label{eq:fidelity}
\end{equation}
where \(d=2^n\) is the Hilbert-space dimension. This is the entanglement
fidelity of the corresponding unitary channels \cite{Horodecki1999GeneralTeleportationChannel}.

To prevent the model from always collapsing to using complex circuits, we introduce a 
regularisation factor $J$ that penalises larger circuits by incorporating 
gate error. For a quantum processor with CZ error rate $p_{\rm CZ}$ and single-qubit gate
error rate $p_{\rm 1Q}$ and a circuit with $n_{\rm CZ}$ CZ gates and $n_{\rm 1Q}$
single-qubit gates, the depolarisation cost is

\begin{equation}
    J(C) = -\left[n_{\rm CZ} \log\left(1-p_{\rm CZ}\right) + n_{\rm 1Q}\log\left(1 - p_{\rm 1Q}\right)\right]
    \label{eq:depol-cost}
\end{equation}

For our experiments, we use $p_{\rm CZ} = 3.3\times10^{-3}$ and 
$p_{\rm 1Q} = 3.5 \times 10^{-4}$, which are the reported CZ and one-qubit error
rates respectively for Google's Willow processor \cite{GoogleQuantumAIWillowSpecSheet2024}.

The final circuit score is then determined by 

\begin{equation}
    \mathcal{S}(C) = F(C, U_{\star}) - J(C)
    \label{eq:circuit-score}
\end{equation}
This ensures that the circuit is rewarded for matching the target unitary but
at the same time penalised for doing it with an excessive implementation. This
score makes the skeleton labels and final circuit selection hardware-aware,
allowing the model to be aware of real world hardware tradeoffs.

\subsection{Training Dataset}
\label{sec:training-data}
To train both models, we use a synthetic dataset pairing Hamiltonians with
circuits and skeletons. The training data generation process is summarised in
Fig.~\ref{fig:main-figure}a. Given a target pair $(H,T)$, we first generate
$N=5000$ random local-gate configurations for each of the seven skeletons in
the template library. The random local-gate configurations used for dataset generation are drawn from
a fixed single-qubit proposal library. This library is produced once by a
separate single-gate diffusion model trained on small $SU(2)$ neighborhoods of
the 24 single-qubit gates of the Clifford group modulo global phase. Here the Clifford group is defined as the normaliser of the Pauli group $\mathcal{P}_1
= \{\pm I,\pm iI,\pm X,\pm iX,\pm Y,\pm iY,\pm Z,\pm iZ\}$ in $\SU(2)$

\begin{equation}
    \mathrm{Cliff}_1
= \{V\in U(2): V\mathcal{P}_1V^\dagger=\mathcal{P}_1\}
\label{eq:clifford-set}
\end{equation}
Thus the initial candidate circuits are not
Haar-random: they are generated from a learned near-Clifford local-gate prior.

For a fixed skeleton $S$, these random configurations are ranked by
their fidelity $F(C)$ to the target unitary $U_\star=\exp(-iHT)$, and the
best $k=8$ candidates are retained. Each retained candidate is then locally refined on the same skeleton for
further optimisation to achieve better fidelity. This refinement is implemented as a short differentiable optimisation
over the local gate parameters. Each local gate is represented by a unit
quaternion $q_j\in S^3\simeq SU(2)$. Starting from the proposed circuit
$\mathbf{q}^{(0)}$, we minimise
\begin{equation}
\mathcal{L}_{\rm ref}(\mathbf{q})
=
1-
F\!\left(C_S(\mathbf{q}), H, T\right)
\end{equation}
by backpropagating through the circuit unitary. After each optimiser step, each
quaternion is projected back to unit norm,
\begin{equation}
q_j \leftarrow \frac{q_j}{\|q_j\|},
\end{equation}
so the refined gates remain valid elements of $SU(2)$. The CZ pattern and
skeleton are fixed throughout. We limit this process to 60 refinement
steps. After this refinement, for each $(H,T,S)$, we keep the refined
candidate with the largest final fidelity. This produces one training example
$(H,T,S)\mapsto C_\star$ for each skeleton $S$ in the library. These examples
are used to train the skeleton-conditioned $SU(2)$ diffusion model.

To train the skeleton selector, we further collapse this dataset across
skeletons. For each target $(H,T)$, we compare the seven refined circuits
$\{C_\star(S)\}_S$. Among skeletons whose refined circuit reaches
$F \geq 0.99$, we choose the skeleton maximizing the
hardware-aware score
\begin{equation}
\mathcal{S}(C_\star(S))
=
F(C_\star(S), H, T)-J_{\rm depol}(C_\star(S)).
\end{equation}
If no skeleton reaches the fidelity threshold, we choose the skeleton whose
refined circuit has the largest fidelity. This produces supervised selector
examples $(H,T)\mapsto S_\star$.

\subsubsection*{Hamiltonian Training Targets}
In these experiments, the training dataset is constructed 
out of 911 three-qubit Hamiltonian targets spanning a 
distribution of target types. 

\begin{enumerate}
    \item Random Pauli Hamiltonians: 512 linear combinations
    of a small pool of Pauli matrix based Hamiltonians
    XII, IZI, IIZ, XXI, IZZ, ZXZ. These give broad
    coverage of generic-looking Hamiltonians.
    \item Local Qubit Operations: 9 basic local operations $\{XII, YII, ZII, IXI, IYI, IZI, IIX, IIY, IIZ\}$ to allow the model to learn
    purely local operations. 
    \item Simple Entangling Operations: $\{XXI, YYI, ZZI\}$
    on qubits 0 and 1 and $\{IXX, IYY, IZZ\}$ on qubits
    1 and 2 to make the model distinguish between the
    two single-CZ circuit skeletons. 
    \item Broader Entanglement Targets: 384 Hamiltonians
    resembling more realistic simulation targets such as
    Heisenberg-XXZ models and transverse-field Ising models.
\end{enumerate}

All targets are implemented with different time ranges $T \sim 0-1$.

\subsection{Skeleton Selector}
\label{subsec:skeleton-selector}
The skeleton selector is trained to predict the best circuit skeleton 
for the given Hamiltonian $H$ and time evolution $T$. This model is 
 implemented as a small multilayer perceptron (MLP). Given 
$H$ and $T$, it outputs logits over the finite skeleton library

\begin{equation}
    p_\psi(S\mid H,T)=\mathrm{softmax}(f_\psi(H,T)).
\end{equation}

The MLP is implemented with two hidden layers of width 256 and 
a SiLU nonlinearity \cite{elfwing2018sigmoid} as the activation function, followed by a linear output
layer with one logit per skeleton. Here, $\psi$ represents the trainable parameters of the MLP. The selector is trained using the 
cross-entropy loss against the training dataset label $S_{\star}$
\begin{equation}
\mathcal{L}_{\rm sel}
=-\log p_\psi(S_\star\mid H,T)
\end{equation}

At inference time we choose the most probable skeleton,
\begin{equation}
\hat S = \arg\max_S p_\psi(S\mid H,T)
\end{equation}
The selector therefore learns the hardware-aware skeleton choice, but
does not itself optimise circuit fidelities or gate parameters.

\subsection{Skeleton-conditioned Diffusion Model}

For a fixed skeleton $S$, the continuous degrees of freedom are the local
single-qubit gates filling its slots. We write a circuit on skeleton $S$ as
$C_S(\mathbf{q})$, where
\begin{equation}
\mathbf{q}=(q_1,\ldots,q_m)\in \SU(2)^m
\end{equation}
is the stack of local gates and $m$ is the number of active local-gate slots.
Different skeletons have different values of $m$, so in implementation all
gate stacks are padded to the maximum slot count, $m_{\max}=18$, and an
active-slot mask indicates which entries belong to the chosen skeleton.

The diffusion model is a skeleton-conditioned token denoiser on this product
manifold with each local gate $q_j$ treated as a token. The noisy gate
quaternion, the slot index, the selected skeleton, and the diffusion timestep
are embedded into a shared hidden space. We also include special conditioning
tokens encoding the target Hamiltonian/evolution time and the selected
skeleton. The resulting sequence has the form
\begin{equation}
[\mathrm{target}(H,T),\, \mathrm{skeleton}(S),\, q_1,\ldots,q_{m_{\max}}],
\end{equation}
and is processed by a Transformer encoder \cite{vaswani2017attention}.

The model outputs one three-dimensional tangent vector for each local-gate
slot,
\begin{equation}
\epsilon_\theta(\mathbf{q}_t, t,H, T, S)
=
(\epsilon_{\theta,1},\ldots,\epsilon_{\theta,m_{\max}}),
\qquad
\epsilon_{\theta,j}\in \mathbb{R}^3 .
\end{equation}
where $t$ represents the diffusion timestep, which is distinct from $T$
the target unitary evolution time interval. 
Inactive padded slots are masked out. For active slots, the target is the
heat-kernel denoising vector derived in Sec.~\ref{sec:diffusion-math}, and the model is trained with
the mean-squared denoising loss
\begin{equation}
\mathcal{L}_{\rm diff}
=
\mathbb{E}
\left[
\sum_{j\in \mathrm{active}(S)}
\left\|
\epsilon_{\theta,j}
-
\epsilon^{\rm target}_{j}
\right\|^2
\right].
\end{equation}
In this work we use a Transformer encoder with hidden width
256, four self-attention layers, four attention heads, and maximum sequence
length corresponding to 18 local-gate slots. This architecture lets the denoising prediction for each gate depend not only
on its own noisy value, but also on the target Hamiltonian, the selected
entangling skeleton, and the other local gates in the circuit.

\subsubsection*{Local Refinement}
The diffusion model is used as a proposal distribution which is then further
refined to obtain a high-fidelity circuit. At inference, one reverse-diffusion trajectory produces one proposed local-gate
stack. In our experiments, we sample a finite batch of proposals for each
target and selected skeleton, drawing $N$ proposals,
rank them by proposal fidelity, and refine the top $k$ candidates for 60
gradient steps by the same process as in Sec.~\ref{sec:training-data}. 

\section{Results}

We now present the results of the main quantum 
circuit synthesis experiments with the above model
architecture. 
All computational details necessary to replicate these results are provided
in the appendix. The code is available in the accompanying repository \cite{singh_su2diffusion_2026}.

\subsection{Unitary Compilation}
\label{sec:unitary-compilation}
We test the fully autonomous inference pipeline consisting of the skeleton
selector and $\SU(2)$ diffusion model on Hamiltonian targets outside the
training set, drawn from five physically motivated Hamiltonian families. For a
given family, we sample 20 Hamiltonian evolution targets $(H,T)$. For each target, the
diffusion model produces 64 proposal circuits on the selected skeleton. The proposals
are ranked by fidelity to the target unitary, and the top 16 undergo local
refinement for 60 optimiser steps. Thus each Hamiltonian family contributes
$20\times 16 = 320$ refined candidate circuits $C$, each scored by its unitary
fidelity with the intended target, $F(C,\exp(-iHT))$. To quantify performance,
we report the fraction of refined candidates in a given Hamiltonian family that
cross the desired fidelity threshold $F\geq 0.99$. We repeat these experiments
across three independent random seed configurations. The Hamiltonian families are:

\begin{enumerate}
    \item TFIM: 
    $H = J(ZZI+IZZ) + h(XII + IXI + IIX)$
    \item Heisenberg-XXZ: $H = J_{xy} (XXI + YYI + IXX + IYY)
    + J_z (ZZI + IZZ)$
    \item Local Pauli: single-qubit targets as described 
    in Sec.~\ref{sec:training-data}.
    \item Mixed Pauli: random mixtures of local, two-body, and selected mixed Pauli terms, with random signs and coefficients.
    \item Near-Threshold Probes: deliberately stronger entangling targets designed to stress the $F \geq 0.99$ decision boundary. We alternate between higher-coupling TFIM-like Hamiltonians and dense nearest-neighbor two-body Pauli Hamiltonians, with longer evolution times than the standard TFIM/XXZ families.
\end{enumerate}

In this experiment, we compare unitary compilation success rate via 
diffusion to two baselines: Haar-random search and 
generated search. All three methods start with the same
circuit skeleton chosen by the selector and undergo the same
60-step local refinement process on their best candidates. The key difference is 
how they produce initial circuit proposals.

The Haar-random proposals are constructed
by filling local-gate slots on the skeleton independently with a gate chosen
randomly from the Haar (uniform) distribution on $\SU(2)$. 64 initial
Haar proposals are generated per target, with the top 16
undergoing refinement. 

The generated search baseline instead 
draws each local gate from the near-Clifford library described 
in Sec.~\ref{sec:training-data}. Generated search is therefore 
a stronger baseline than Haar-random because it samples proposals using near-Clifford
structured but still target-agnostic local-gate prior. As
with the Haar-random baseline, this method also populates 
gates independently. For generated search, we generate
2000 random candidate circuits and refine the top 16 
for 60 refinement steps. Because generated search uses a structured local-gate prior and is allowed many
more raw candidates than diffusion, it provides a relatively strong
target-agnostic baseline.

\begin{figure*}
    \centering
    \includegraphics[width=1.0\linewidth]{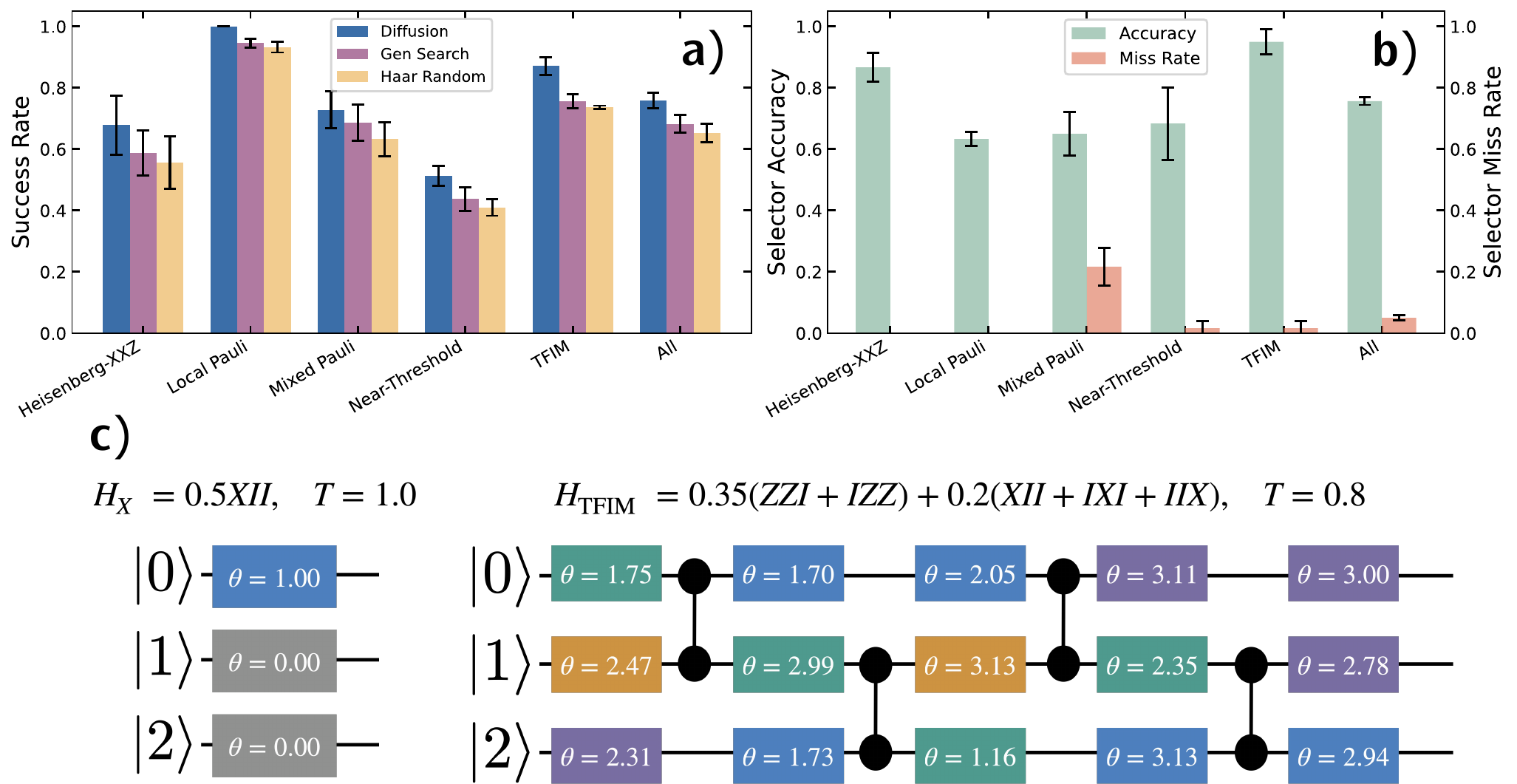}
    \caption{Unitary Compilation. 
a) Post-refinement success rates on five held-out Hamiltonian families, with 20 targets per family and three independent random seed configurations. For each target, all methods use the same selector-chosen skeleton and refine the top candidate circuits under the same local optimisation protocol. The Hamiltonian-conditioned \(\mathrm{SU}(2)\) diffusion model consistently outperforms both generated-search and Haar-random baselines. 
b) Skeleton-selector diagnostics on the same targets. Selector accuracy is the fraction of targets for which the learned selector matches the cost-aware oracle skeleton
choice. Miss rate is the fraction of targets for which some skeleton produces a successful circuit with \(F\geq 0.99\), but the selector-chosen skeleton produces none. Error bars show variability across random seed configurations. (c) Representative circuits produced by the full pipeline. The local
\(H=0.5XII\), \(T=1.0\) target is compiled to a 0-CZ circuit with a single
\(R_x(1.0)\) rotation, while the TFIM target is compiled to a 4-CZ
entangling circuit. Coloured boxes denote single-qubit rotations.  Colour encodes
the dominant component of the qubit rotation axis for visualisation purposes: blue for $X$, gold for $Y$, and green for $Z$. Purple denotes rotation axes where no one direction dominates and grey denotes zero rotations, i.e. the identity operation. The printed number $\theta$ is the gate rotation
angle around the selected rotation axis in radians.}
    \label{fig:success-rate}
\end{figure*}
 
Fig.~\ref{fig:success-rate}(a) shows the post-refinement success rate for each
held-out target family. The Hamiltonian-conditioned diffusion model outperforms
both baselines across all families. As expected, generated search performs better
than Haar-random sampling, reflecting the additional structure of its learned
single-qubit proposal library. The highest success rates occur for the Local
Pauli family, which consists of non-entangling single-qubit targets. Diffusion
reaches \(100\%\) success on this family which indicates that the model robustly
generalises to these simple local evolutions. The most challenging family is the
Near-Threshold set, which was deliberately constructed to stress the
\(F\geq 0.99\) decision boundary. Even there, diffusion reaches \(51\%\) success,
exceeding both baselines. Aggregated over all five families and three random seed
configurations, diffusion achieves \(75.8\pm 2.6\%\) success, compared with
\(68.3\pm 2.8\%\) for generated search and \(65.3\pm 3.0\%\) for Haar-random
sampling.

To evaluate the selector portion of the model, we define selector accuracy and
selector miss rate, shown in Fig.~\ref{fig:success-rate}(b). For each held-out
target \((H,T)\), we compare the selector's chosen skeleton \(S\) to a post-hoc
oracle skeleton \(S_\star\). The oracle is obtained by running the diffusion model
on every skeleton in the template library, refining the top candidates for each
skeleton, and then applying the same thresholded hardware-aware scoring rule used
to construct selector labels in the training set. The selector accuracy is the
fraction of targets in a family for which \(S=S_\star\). We also define the miss
rate as the fraction of targets for which at least one skeleton produced a
refined candidate with \(F\geq 0.99\), but the selector-chosen skeleton produced
none.

On all families the selector accuracy is much higher than the miss rate. Overall the
selector accuracy is $75.6 \pm 1.2\%$ while the overall miss rate is $5.0 \pm 0.8\%$.
The miss rate is highest on the Mixed Pauli family at $21.7 \pm 6.2\%$. This is consistent with the
heterogeneous structure of these targets: depending on the random signs and
coefficients, individual instances can range from nearly local to genuinely
entangling. As a result, low-cost local skeletons can sometimes achieve high but
sub-threshold fidelities, while an entangling skeleton is required to cross
\(F\geq0.99\). The mixed family therefore exposes the sharpest selector boundary
between inexpensive near-solutions and successful entangling decompositions.
Fig.~\ref{fig:success-rate}(c) gives a concrete view of what the reported
success rates correspond to at the circuit level. For the local
\(H=0.5XII\) target, the full pipeline reduces to the expected 0-CZ solution:
a single approximately \(R_x(1.0)\) rotation on qubit 0 and near-identity
rotations on the other qubits. This provides a useful sanity check that the
selector does not introduce unnecessary entangling gates. For the TFIM target,
the same pipeline instead selects an entangling 4-CZ skeleton and fills it with
nontrivial local rotations, producing a concrete circuit whose unitary matches
the Hamiltonian evolution above the \(F\geq0.99\) threshold. 

\subsection{Model Ablations}
We next evaluate how much each element of the pipeline contributes to the overall
synthesis success. To do this, we run an ablation version of the unitary compilation experiment on the same suite of \(20\times 5=100\) Hamiltonian targets. Instead
of reporting success family-by-family, we aggregate over all targets and compare
the overall success rate of several pipeline configurations. The ablation
experiments are repeated over the same three seed configurations as in
Sec.~\ref{sec:unitary-compilation}. The configurations are summarised in
Table~\ref{tab:ablation-configurations}.

For each target, the diffusion and Haar configurations generate 64 raw proposals,
while generated search generates 2000 raw proposals from the learned single-qubit Clifford
library. In all refinement-based configurations, the top 16 proposals per target
are refined for 60 optimisation steps and included in the success statistics. The
``diffusion only'' configuration uses the same top 16 diffusion proposals but
does not apply local refinement. For each configuration, the
success rate is defined as the fraction of
$16$ top proposals across all the $100$ targets that achieve the $F\geq0.99$
threshold. 

\begin{table*}[t]
\centering
\begin{tabular}{llll}
\hline
\textbf{Configuration} &
\textbf{Skeleton} &
\textbf{Circuit Proposal} &
\textbf{Refinement} \\
\hline

Haar + Refinement &
Learned Selector &
Haar-Random SU(2) gates &
Yes \\

Generated-Search + Refinement &
Learned Selector &
Local-Clifford Gate Search &
Yes \\

Full Model (Diffusion + Refinement + Selector) &
Learned Selector &
Hamiltonian-conditioned Diffusion &
Yes \\

Diffusion + Refinement + Fixed Skeleton &
4-CZ &
Hamiltonian-conditioned Diffusion &
Yes \\

Diffusion + Refinement + Oracle Skeleton &
Oracle Skeleton &
Hamiltonian-conditioned Diffusion &
Yes \\

Diffusion Only &
Learned Selector &
Hamiltonian-conditioned Diffusion &
No \\

\hline
\end{tabular}
\caption{
Summary of circuit-synthesis configurations used in the ablation study.
Each configuration differs in how the circuit skeleton is selected, how the
initial gate parameters are proposed, and whether continuous refinement is applied.
}
\label{tab:ablation-configurations}
\end{table*}

\begin{figure*}
    \centering
    \includegraphics[width=1.0\linewidth]{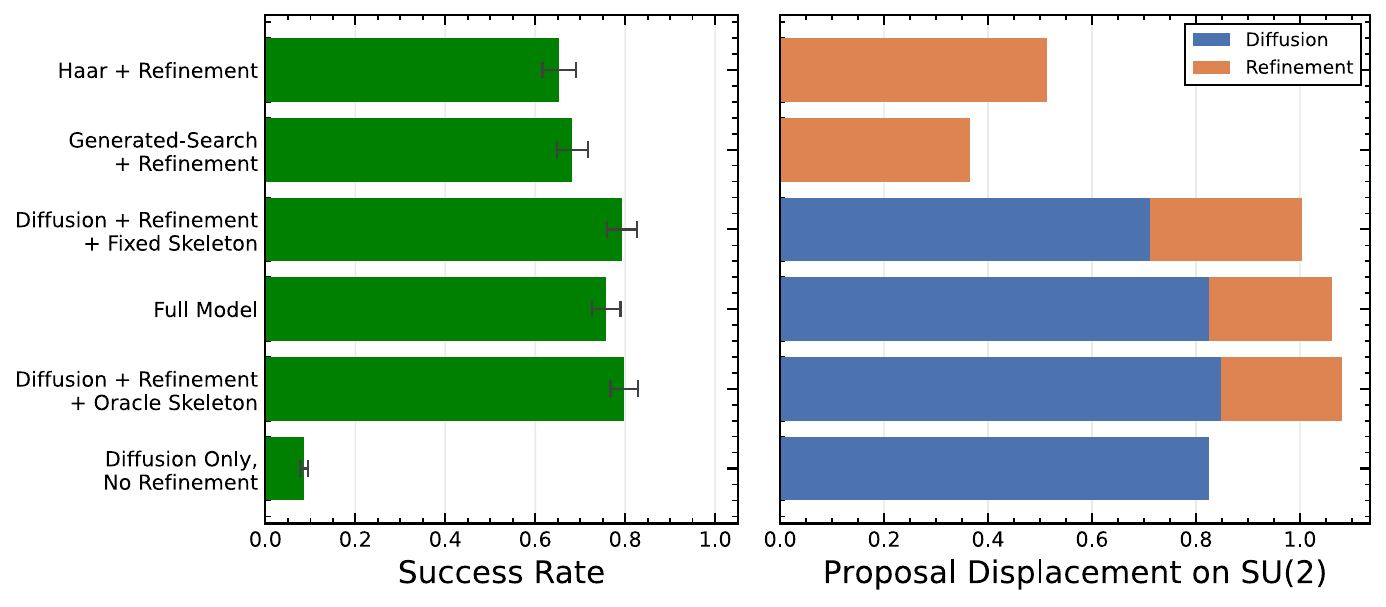}
    \caption{Results of the ablation experiments. (a) Success rate vs
    configuration. The success rate is defined as the fraction of 
    final proposals that have a fidelity $F\geq 0.99$. (b) Geometric
    displacement $d_{\rm circuit}$ of circuit proposals on $\SU(2)$ as defined by Eq.~\ref{eq:circuit-geodesic-distance}. Blue
    represents the displacement during the diffusion process
    from Haar-random to non-refined proposals. Orange
    represents displacement by local refinement.}
    \label{fig:ablation-success-and-distance}
\end{figure*}

The results are shown in Fig.~\ref{fig:ablation-success-and-distance}(a).
The full model configuration outperforms the Haar/Generated-Search +
Refinement configurations, supporting the previous finding that
diffusion beats both baselines. More interestingly, the full model is 
beaten marginally by replacing the local skeleton selector with 
a fixed skeleton (in this case 4-CZ) and by the brute force
oracle skeleton selector used to create the training dataset. 
The full model achieves $75.8 \pm 3.1\%$ while the fixed skeleton
achieves $79.3 \pm 3.3\%$ and the oracle skeleton achieves
$79.8 \pm 3.0\%$. This suggests that the learned selector is 
marginally limiting the performance of the overall circuit 
synthesis. 

Diffusion without refinement has the worst success
rate at $8.6\pm 0.8\%$. This suggests that local refinement is crucial to the circuit
synthesis process, although local refinement by itself starting from a Haar-random
start clearly underperforms diffusion and local refinement combined. One can interpret
this geometrically by considering the displacement of circuit proposals on $\SU(2)$ during the synthesis process.
For two elements $A, B \in \SU(2)$, the geodesic distance between them is

\begin{equation}
    d_{\rm SU(2)}(A,B) = \arccos\left(\frac{1}{2}\left|\mathrm{Tr} (A^{\dagger}B)\right|\right)
    \label{eq:geodesic-distance}
\end{equation}

We generalise this to circuits containing $m$ elements from $\SU(2)$. If
$Q = (q_1, \dots, q_m)$ and $R = (r_1,\dots, r_m)$ are circuits, then the circuit distance
is defined as

\begin{equation}
    d_{\rm circuit}(Q,R) = \frac{1}{m}\sum_{j=1}^m d_{\rm SU(2)}(q_j, r_j)
    \label{eq:circuit-geodesic-distance}
\end{equation}

We focus on the geodesic displacement instead of a metric like total 
distance traveled as we are not interested in proposal trajectories that
for example oscillate locally without moving to a higher fidelity solution.
In Fig.~\ref{fig:ablation-success-and-distance}(b), we plot the average circuit
displacement between the starting points and endpoints for all the configurations. For
the configurations without diffusion, this corresponds to the displacement under
local refinement. For the configurations with diffusion and local refinement, 
this has a component from the diffusion step and from the refinement step. 
We see that all diffusion configurations 
demonstrate the same circuit displacement, but the local refinement configurations
undergo a smaller displacement that allows much better performance. Thus we
see that diffusion and local optimisation are complementary: diffusion 
moves the proposal into a beneficial starting point for refinement,
which is a much more local operation. 
 
\section{Controllable Solution Style: Angle Steering}

To test whether the diffusion prior controls not only target fidelity but also
the style of synthesised circuits, we train two otherwise identical
skeleton-conditioned diffusion models on biased solution datasets. In the
experiments above, each \((H,T,S)\) training example kept the
refined circuit with the highest final fidelity. Here, for the same target and
skeleton libraries, we instead construct two datasets: one retaining the
successful refined circuit with the lowest total local rotation angle, and one
retaining the successful refined circuit with the highest total local rotation
angle. The skeleton selector is trained once on the combined low- and high-angle datasets, giving a shared discrete skeleton policy, while the two diffusion priors are trained separately on the corresponding low-angle and high-angle circuit parts.

For a circuit \(C\) with local \(\SU(2)\) gates represented by quaternions
\(q_j=(w_j,\vec v_j)\), we define the total local rotation angle as
\begin{equation}
    A(C)=\sum_j 2\arccos(|w_j|),
    \label{eq:total-local-angle}
\end{equation}
where the sum is over active local-gate slots. We then generate circuit
proposals for representative Hamiltonian targets, refine them, and retain only
successful circuits satisfying \(F\geq0.99\). For each successful refined
circuit, we compute \(A(C)\).

\begin{figure}
    \centering
    \includegraphics[width=1.0\linewidth]{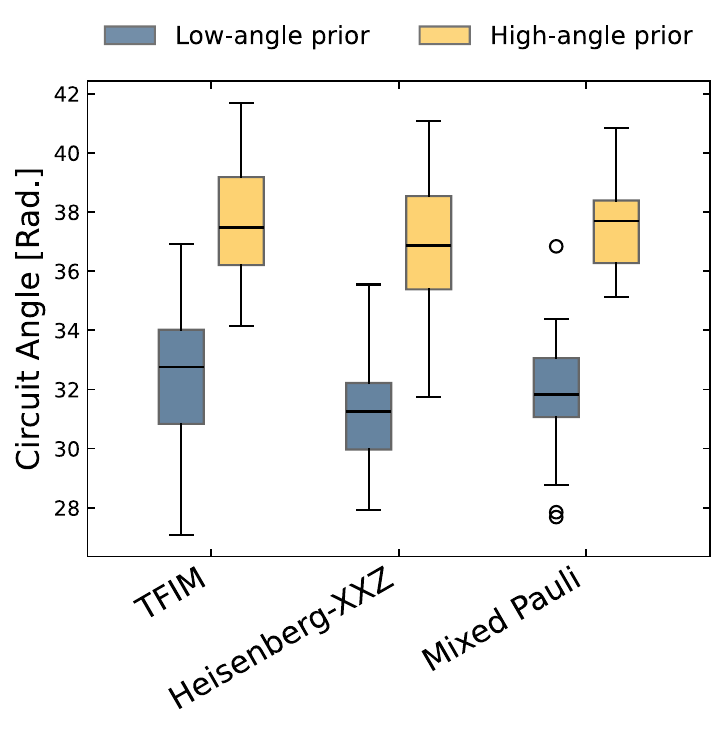}
    \caption{Distribution of total local rotation angle for successful refined circuits generated by low-angle and high-angle diffusion priors. Only circuits satisfying $F\geq0.99$ are included. Across all three entangling target families, the high-angle prior shifts the successful solution distribution toward larger total local rotation angle. This shows that the skeleton-conditioned diffusion prior can steer continuous circuit style on $\mathrm{SU}(2)^n$, rather than only optimizing final fidelity.}
    \label{fig:angle-distribution}
\end{figure}

As shown in Fig.~\ref{fig:angle-distribution}, the high-angle prior consistently
produces successful circuits with larger total local rotation angle than the
low-angle prior across target Hamiltonians. This indicates that
the learned diffusion prior captures controllable structure in the continuous
\(\SU(2)^n\) degrees of freedom, while the selector handles the discrete
entangling-template choice.

\section{Hardware-Aware Fidelity-Complexity Frontiers}
Because the skeleton-conditioned diffusion model can generate local-gate
proposals for any skeleton in the template library, we can inspect the
fidelity-cost landscape that the learned selector is meant to approximate. For a
fixed target Hamiltonian evolution, we run the diffusion model across all
skeletons, refine the resulting proposals, and compare the selector's chosen
skeleton with the empirical distribution of refined fidelity
 across the template library.

\begin{figure*}
    \centering
    \includegraphics[width=1.0\linewidth]{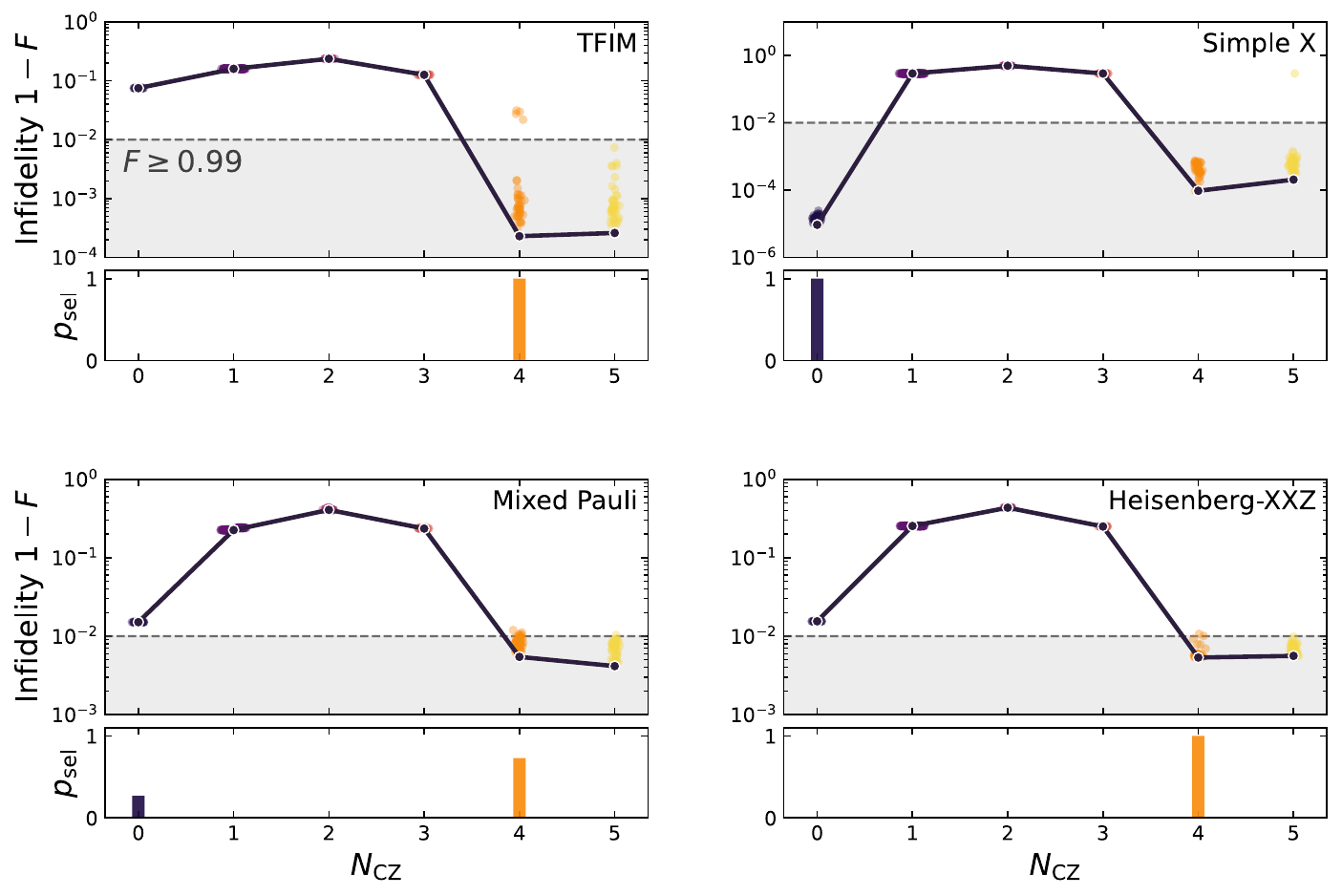}
    \caption{Circuit Depth-Fidelity frontier of learned skeleton-conditioned circuit proposals.
For representative held-out targets, we sample and refine candidates on each
skeleton and plot the best infidelity \(1-F\) obtained at each CZ count
\(N_{\rm CZ}\). The shaded region indicates the success threshold
\(F\geq 0.99\). Lower panels show the learned selector probability
\(p_{\rm sel}=p(S\mid H,T)\). The local Simple-X target is solved by the
0-CZ skeleton, whereas the entangling TFIM target shows a
sharp transition at four CZ gates. The Mixed Pauli and Heisenberg-XXZ targets illustrate an
intermediate case: a cheap local skeleton approaches the target but does not
cross the success threshold, while the 4-CZ skeleton succeeds. Overall, the
selector concentrates on the first successful complexity class, supporting the
interpretation that it learns a hardware-aware skeleton choice rather than
simply selecting the deepest template.}
    \label{fig:pareto-frontiers}
\end{figure*}

Fig.~\ref{fig:pareto-frontiers} shows this fidelity--complexity landscape for
four representative targets: TFIM, Simple X, Mixed Pauli, and
Heisenberg-XXZ. The Simple X target is a local Hamiltonian evolution generated
by a single Pauli \(X\) term on one qubit, and is correctly solved by the
0-CZ skeleton. By contrast, the TFIM and Heisenberg-XXZ targets show a sharp
complexity transition: skeletons with fewer than four CZ gates fail to reach
the \(F\geq0.99\) threshold, while the 4-CZ skeleton succeeds. The selector
places essentially all probability on this first successful complexity class,
rather than defaulting to the deepest 5-CZ skeleton. 
The Mixed Pauli target illustrates a more ambiguous boundary case. A local
0-CZ circuit reaches relatively high fidelity but remains just below the
success threshold, reflecting the substantial local component of this target. The nonzero selector mass
on the 0-CZ skeleton is interpretable as uncertainty near a
local-to-entangling complexity boundary.
Both the 4-CZ and 5-CZ skeletons cross the threshold and despite the fact 
that 5-CZ 
achieves the best fidelity, the selector still assigns its
largest probability to the 4-CZ skeleton. This reflects the tradeoff between 
circuit depth and fidelity encoded through the hardware-aware score during training.

\section{Discussion and Outlook}
In this work we have demonstrated a new approach to 
quantum circuit synthesis using diffusion models. 
By breaking the problem up into circuit
template selector
and diffusion gate prior generation components, this work
naturally integrates physical structure and design 
constraints into the generation process. This work
also offers an interpretable machine learning approach: the circuit template selection is a direct function
of the tradeoff between final fidelity and hardware 
cost while the final gate configuration now has a
geometric interpretation inherited naturally from
the structure of $\SU(2)$. 

An important finding is that diffusion serves as the 
circuit proposal \textit{prior} that is then refined
into a high fidelity solution by local optimisation. 
The model ablation results clearly demonstrate that
diffusion alone has a weak success rate, but crucially
so does refinement alone from a Haar-random initial start.
The strongest results were in the ablations that combined
diffusion with local optimisation. Thus
diffusion's key role is as a learned proposal prior
on $\SU(2)$ which then allows local optimisation 
to achieve the required fidelity. Fig.~\ref{fig:ablation-success-and-distance}(b) demonstrates that this optimisation is indeed local, as the displacement on
$\SU(2)$ under the optimisation process is a smaller
perturbation on the large overall displacement
under the diffusion process. 

In Fig.~\ref{fig:ablation-success-and-distance}(a),
we also see that the full model is marginally beaten
by diffusion and refinement with a fixed 4-CZ 
circuit skeleton aggregated across all targets. This reflects the impact of the circuit
selector component. The 
diffusion component maximises fidelity 
given a target Hamiltonian evolution and circuit template, but is agnostic to hardware penalties. The
selector by contrast encodes the hardware penalty 
and can reject deeper circuits that achieve higher
fidelity at greater hardware cost. This is shown 
notably in the Fig.~\ref{fig:pareto-frontiers} Mixed
Pauli case where the 4-CZ template is selected despite
5-CZ having better fidelity. Therefore
the selector component implements the hardware aware
design policy that can be customised to a particular
QPU architecture. 
A key advantage of generative modeling is the ability
to create a distribution of solutions, and the present
work demonstrates this in the angle steering 
experiments. By biasing the model pipeline's training
data, we are able to create circuits with large and
small rotation angles for the same Hamiltonian 
targets as shown in Fig.~\ref{fig:angle-distribution}. 
This can be generalised to other properties tailored
to the particular quantum hardware. In future work, the same strategy could be used to favor circuits with different axis compositions, lower pulse-duration proxies tailored
to the hardware available, reduced sensitivity to gate perturbations, or lower device-calibrated error costs.

This work has focused on three-qubit unitary compilation as three
qubits offer a nontrivial testbed for circuit compilation that 
has been used in previous efforts too \cite{furrutter2024quantum}.
Nevertheless, a clear extension is extending this to multiple
qubits and scaling up the corresponding circuit templates as well. Diffusion models have been scaled up considerably in other applications \cite{podell2023sdxl} and this can be exploited to scale
up the present diffusion architecture to larger numbers of qubits.

As the number of qubits scales up, the circuit templates will scale
up too and should reflect more complex connectivity graphs such
as nearest-neighbour on a 2D superconducting qubit grid \cite{GoogleQuantumAIWillowSpecSheet2024}, IBM's superconducting
heavy-hex architecture \cite{hetenyi2024heavyhex}, or all-to-all 
connectivity as demonstrated in recent trapped-ion QPUs \cite{ransford2026alltoall}. The reprogrammability of neutral 
atom connectivity graphs \cite{bluvstein2024logical} also
offers an interesting test case for the modular selector-diffusion 
architecture as different selector submodels can potentially be trained for 
different QPU configurations without having to retrain the gate generation diffusion component. As the number of qubits scales up,
the circuit selector could transition to a learned mechanism
for proposing hardware-aware templates rather than learning
to select from
a pre-initialised bank. 

Another extension is to modify the hardware cost to reflect
real hardware more accurately. The hardware cost used here is a simple depolarising proxy based only on one- and two-qubit gate error rates. Real devices demonstrate richer and more structured error
models including features such as crosstalk and qubit leakage errors \cite{miao2023overcoming}.
In this work we have introduced a generative
approach to quantum circuit synthesis that incorporates
both the physical geometry of the problem and constraints
of real-world hardware. The method separates the synthesis problem into a discrete circuit-skeleton choice and a continuous local-gate generation problem. A learned selector chooses an entangling CZ template using a hardware-aware fidelity-cost objective, while a conditioned diffusion model generates local one-qubit gates directly on the Lie group \(\mathrm{SU}(2)^n\).

Our results suggest that quantum circuit synthesis can be usefully viewed as conditional generation on a hybrid design space: a discrete entangling skeleton together with continuous local gates on a compact Lie group. 
Overall, this work shows that respecting both the geometry of quantum gates and the hardware cost of entanglement can make diffusion models useful not merely as random circuit generators, but as structured proposal priors for hardware-aware quantum compilation.

\vspace{3mm}

\begin{acknowledgments}
The author would like to thank Dhruv Devulapalli and Ali Malik 
for fruitful discussions. OpenAI's GPT-5.5 assisted in the execution
of the experiments in this work. 
\end{acknowledgments}

\appendix

\section{Computational Details}

All experiments in the main text were carried out for three-qubit target
unitaries of the form \(U_\star=\exp(-iHT)\). 
The skeleton selector is a three-layer multilayer perceptron with hidden
dimension 256 and SiLU activations. It was trained for 4000 optimization steps
using cross-entropy loss and the AdamW optimiser \cite{loshchilov2019decoupled} with learning rate \(10^{-3}\) and weight decay
\(10^{-4}\). 
The skeleton-conditioned diffusion model is a token transformer operating on
local-gate slots. Each local gate is represented by a unit quaternion, and each
slot token received a learned slot embedding, template embedding, active-slot
embedding, diffusion-time embedding, and Hamiltonian target conditioning. The
transformer used hidden dimension 256, 4 encoder layers, 4 attention heads, GELU
feed-forward blocks \cite{hendrycks2016gelu} with multiplier 4, dropout 0, and a final MLP head predicting
a three-dimensional Lie-algebra noise vector for each active local-gate slot.
The diffusion schedule used 100 steps with a linear
\(\beta\)-schedule from \(10^{-4}\) to \(5\times10^{-3}\). The heat-kernel
epsilon target used 64 spectral terms. The model was trained for 2000 AdamW
steps with batch size 256, learning rate \(2\times10^{-4}\), and weight decay
\(10^{-4}\).

The experiments were performed on
a single NVIDIA A100 GPU. One full training-and-evaluation experimental run took approximately
6--7 hours, including dataset construction, selector training, diffusion-model
training, and evaluation.

\bibliography{references}

\end{document}